\documentclass{iopart}

\usepackage{amssymb}

 \newtheorem{Th}{Theorem}
  \newtheorem{Co}{Corollary}

\def\be{\begin{equation}}
\def\ee{\end{equation}}
\def\bea{\begin{eqnarray}}
\def\eea{\end{eqnarray}}

\def\p{\partial}
\def\a{\alpha}

\def\d{\delta}

\def\k{\kappa}
\def\lb{\lambda}

\def\T{\Theta}

\def\vfi{\varphi}

\def\ov{\overline}

\def\l2{{$L^2(-\pi ,\pi )$ }}

\newcommand{\Sc}{Schr\"odinger }

\begin{document}

 \large

\title{SUSY transformation of the Green function and a trace formula}

\author{B F Samsonov$^{\dag}$, C V Sukumar$^{\S}$ and A M Pupasov$^{\dag}$}

\address{\dag\ Physics Department of Tomsk State
 University, 36 lenin Avenue, 634050 Tomsk, Russia}

\address{\S \,Wadham College, Oxford OX1 3PN, England}

\ead{\mailto{samsonov@phys.tsu.ru},
\mailto{c.sukumar1@physics.oxford.ac.uk}
 }

%\baselineskip=18pt
%\maketitle

\begin{abstract}
\baselineskip=16pt
\noindent
An integral relation is established between the
Green functions corresponding to
two Hamiltonians which are supersymmetric (SUSY)
 partners and in general may
possess both discrete and continuous spectra.
It is shown that when the continuous spectrum is present the
trace of the difference of the Green functions for SUSY partners
is a finite quantity which may or may not be equal to zero
despite the divergence of the traces of each Green function.
Our findings are illustrated by using the free particle example
considered both on the whole real line and on a half line.
\end{abstract}

\submitto{\JPA}
 %{\small{\today}}

%%%%%%%%%%%%%%%%%%%%%%%%%%%%%%%%%%%%%%%%%%%%%%%%%%%%%%%%%%%%%%%%%%%%%%%%%
%%%%%%%%%%%%%%%%%%%%%%%%%% INTRODUCTION %%%%%%%%%%%%%%%%%%%%%%%%%%%%%%%%%

\section{Introduction}

At present there is a growing interest in the study
of different properties transformations induced by  supersymmetry (SUSY) in Quantum
Mechanics. Recently a special issue of Journal of
Physics A (see vol. 37, No~43, 2004) was devoted to research work in this subject. Despite of the growing
number of papers in this field many questions still remain open and
require further study.
 In particular, the authors are aware of only one paper
\cite{Sukumar}
devoted to the study SUSY transformations at the level of Green
functions. For the case of a transformation deleting the ground state
of the initial Hamiltonian, Sukumar has studied an integral relation
between the Green functions for SUSY partners
 and has formulated conditions leading to the vanishing of some matrix elements of
 a Hamiltonian and related this property to a hidden
 supersymmetry of the system. Transformation of Green functions is
 not explicitly discussed in that paper. Moreover, we have found that
 formula (28) of \cite{Sukumar} relating integrals over Green functions
for SUSY partners may need to be corrected if a continuous
 spectrum is present.

 In this paper we give a simple formula for the Green function
 of the SUSY partner Hamiltonian both for confining and for scattering potentials
  and generalize results of the
 paper \cite{Sukumar} to the case where the continuous spectrum is
 present. As an application of this general formula
 we consider the case of the \Sc
 equation with a scattering potential defined
 both on the whole real axis and on a half line when
 the \Sc equation is reduced to a singular Sturm-Liouville problem.
 Regular
 Sturm-Liouville problem is considered in a separate publication
 \cite{SPphysica}.

 \section{Green function of the \Sc equation}

 In this section we cite some properties of the Green function
 of the one-dimensional \Sc equation for a spectral problem on
 the whole real line (see e.g. \cite{Morse,Levitan0}) which are useful.

 We consider the \Sc equation
 \be\label{se}
(h_0-E)\psi =0 \qquad\qquad h_0=-d^2/dx^2+V_0(x)
\qquad\qquad x\in (a,b)
\ee
supplemented by the boundary conditions $\psi(a)=\psi(b)=0$.
We will concentrate mostly on two cases; these are the whole real
line $a=-\infty$ and $b=\infty$ and  the half line $a=0$ and
$b=\infty$.
We assume that the spectral set $\mbox{spec}h_0$ of
 this problem consists of $M$ discrete  points
with the possibilities $M=0$ or $M=\infty$ and
possibly a continuum part filling the positive semiaxis.

 The definition of the Green
 function used by different authors
\cite{Morse,Levitan0,BerLev}
  may differ by a constant factor. We use a definition of the Green
function represented as the
  kernel of the operator $(h_0-E)^{-1}$
as an operator defined in the corresponding Hilbert space.
 It is well defined for all
$E\notin \mbox{spec}h_0$.
% where $\mbox{spec}h_0$ is the spectrum of
%eigenvalues of $h_0$.
It has two different but equivalent
representations. The first representation is obtained with the help of two
real solutions of equation (\ref{se}) with a fixed value
of the parameter $E\notin \mbox{spec}h_0$, $f_{l0}$ and $f_{r0}$
(``left" and ``right" solutions),
satisfying zero boundary conditions: $f_{l0}(a)=0$,
$f_{r0}(b)=0$. Since they correspond to the same $E$ their
Wronskian $W_0=W(f_{r0},f_{l0})$ does not depend on $x$ and is a function of $E$ only,
and the Green function is
\bea\label{Glr}
& G_0(x,y,E)=f_{l0}(x,E)f_{r0}(y,E)/W_0 \qquad\qquad  x\le y \\
& G_0(y,x,E)=G_0(x,y,E)\,.
\label{Glr2}
\eea
These formulae are clearly equivalent to
\be\label{Gteta}
G_0(x,y,E)=[f_{l0}(x,E)f_{r0}(y,E)\T(y-x)+
f_{l0}(y,E)f_{r0}(x,E)\T(x-y)]/W_0
\ee
where $\T$ is the Heaviside step function.

If the operator $h_0$ is essentially self-adjoint the set of its
discrete spectrum (if present)
 eigenfunctions $\{\psi_n\}$, $n=0,1,\ldots,M$,
$\langle\psi_n|\psi_m\rangle=\d_{nm}$
together with the continuous spectrum eigenfunctions
(also if present) $\psi_k$, $E=k^2>0$,
$\langle\psi_k|\psi_{k'}\rangle=\d(k-k')$,
$\langle\psi_n|\psi_{k}\rangle=0$
 is complete in the Hilbert space
\[
\sum_{n=0}^M\psi_n(x)\psi_n^*(y)+
\int dk\,\psi_k(x)\psi_k^*(y)=\delta(x-y)
\]
 and
the second representation of the Green function may be found in terms of this set as
follows:
\be\label{G0}
G_0(x,y,E)=\sum_{n=0}^M\frac{\psi_n(x)\psi_n^*(y)}{E_n-E}
+\int \frac{\psi_{k}(x)\psi_{k}^*(y)}{k^2-E}dk\,.
\ee
For the spectral problem on the whole real axis the continuous
spectrum is two-fold degenerate and the integrals over $k$ run from
minus infinity to infinity and for the problem on a half line they
run from zero to infinity.

\section{SUSY transformation of the Green function}

It is well-known (see e.g. \cite{Sukumar1}) that there exist three
kinds of SUSY transformations:\\
({\bf i}) deleting the ground state
level of $h_0$\\
({\bf ii}) creating a new ground state level and
\\({\bf iii})
purely isospectral transformation. \\
 In all cases the partner Hamiltonian $h_1=-d^2/dx^2+V_1$ for
$h_0$ is defined by the potential
\be
V_1(x)=V_0(x)-2w'(x) \qquad \qquad  w(x)=[\log u(x)]'
\ee
where $u$ is a real solution to the initial equation $(h_0-\a)u=0$ with
$\a$ known as the factorization constant. We adopt the notation that a
derivative with respect to $x$ is denoted by the prime symbol.
To provide a nonsingular
potential difference $\a$ should be less than or equal to the ground
state energy of $h_0$ if it has a discrete spectrum or lower than the
continuum threshold otherwise.
The functions $\vfi_n=L\psi_n$, $n=1,2,\ldots, M$ describe (unnormalized)
bound states and  $\vfi_E=L\psi_E$ correspond to
(unnormalized) scattering states of $h_1$.
Here
\be\label{L}
L=-d/dx+w(x)
\ee
is
the transformation operator (intertwiner) satisfying
$Lh_0=h_1L$ and $Lu=0$.
The normalization constants are easily obtained with the help of
the factorization property $L^+L=h_0-\a$ where $L^+=d/dx+w(x)$.
The functions
\be\label{chi}
\chi_n=(E_n-\a)^{-1/2}L\psi_n\qquad \qquad
\chi_E=(E-\a)^{-1/2}L\psi_E
\ee
form an orthonormal set.
\begin{Th}
Let $G_0(x,y,E)$ be the Green function for $h_0$. Then for all
three cases enumerated above the Green function for $h_1$ is
\be\label{G1}
G_1(x,y,E)=\frac{1}{E-\a}
[L_xL_yG_0(x,y,E)-\delta(x-y)]\,.
\ee
In case (ii) it has a simple pole at $E=\a$.
In cases (i) and (iii) it is regular at $E=\a$ and can be
calculated as follows:
\be\label{G1a}
G_1(x,y,\a)=\left[
L_xL_y\frac{\partial G_0(x,y,E)}{\partial E}
\right]_{E=\a}.
\ee
Here $L_x$ is the operator given in (\ref{L}) and $L_y$ is the
same operator where $x$ is replaced by $y$.
\end{Th}

{\em Proof.}
In case (i) $u=\psi_0$ and the set
$\{\chi_E,\chi_n, n=1,2,\ldots,M\}$ is complete.
Therefore
\be\label{G1i}
G_1(x,y,E)=
\sum_{n=1}^M\frac{\chi_n(x)\chi_n^*(y)}{E_n-E}
+\int \frac{\chi_{k}(x)\chi_{k}^*(y)}{k^2-E}dk\,.
\ee
Now we replace $\chi$ using (\ref{chi}) which yields
\[
\begin{array}{l}
G_1(x,y,E)=\\
\frac{1}{\a-E}L_xL_y
\left(
\sum_{n=1}^M
[\frac{1}{E_n-\a}-\frac{1}{E_n-E}]\psi_n(x)\psi_n^*(y)+
\int\!
dk[\frac{1}{k^2-\a}-\frac{1}{k^2-E}]\psi_{k}(x)\psi_{k}^*(y)\
\hspace{-.5em}\right)\!.
\end{array}
\]
The statement for $E\ne\a$ follows from here if in the first sum and in the
first integral we express $L\psi$ in terms of $\chi$, make use of the
completeness condition for the set $\chi$ and formula
(\ref{G0}) for $G_0$. The fact that here the sum starts from
$n=1$ and in (\ref{G0}) it starts from $n=0$ cannot cause any
problems since $L\psi_0=0$.
For $E=\a$ formula (\ref{G1i}) can be written in the form
\[
G_1(x,y,\a)=\left[\frac{\p}{\p E}L_xL_y
\left(
\sum_{n=0}^M\frac{\psi_n(x)\psi_n^*(y)}{E_n-E}
+\int \frac{\psi_{k}(x)\psi_{k}^*(y)}{k^2-E}dk
\right)\right]_{E=\a}
\]
from which (\ref{G1a}) follows in this case.

In case (ii) let $\chi_\a\sim 1/u$ be the normalized
 ground state function of $h_1$ corresponding to the new
 discrete level $E=\a$.
  Then
\be\label{G1ii}
G_1(x,y,E)=
\sum_{n=0}^M\frac{\chi_n(x)\chi_n^*(y)}{E_n-E}
+\frac{\chi_\a(x)\chi_\a^*(y)}{\a-E}
+\int \frac{\chi_{k}(x)\chi_{k}^*(y)}{k^2-E}dk\,.
\ee
Now the use of exactly the same transformations as in case (i) reduces (\ref{G1ii})
to (\ref{G1}).

In case (iii) we start from the same formula (\ref{G1i}) with
the only difference that the sum now starts from $n=0$
and following the same line of reasoning as in the earlier cases we get formula (\ref{G1}).
It is interesting to notice the intermediate result
\be\label{G1iii}
G_1(x,y,E)=
\frac{1}{\a-E}L_xL_y[G_0(x,y,\a)-G_0(x,y,E)]
\ee
which makes clear how formula (\ref{G1a}) arises for this case by taking
 the limit $E\to\a$.
The fact that in case (ii) the function (\ref{G1}) has a simple
pole at $E=\a$ is a consequence of the equivalence between
(\ref{G1ii}) and (\ref{G1}).
\hfill$\square$

\begin{Co}
In terms of the special solutions $f_{l0}$ and $f_{r0}$
of the \Sc equation for $h_0$
the Green function $G_1$
for all three cases listed above
 may be expressed as follows:
\bea\label{G1lr}
&G_1(x,y,E)=        \nonumber
\frac{1}{(E-\a)W_0}
[\T(y-x)L_xf_{l0}(x,E)L_yf_{r0}(y,E)\\
&\hphantom{G_1(x,y,E)=}+\T(x-y)L_yf_{l0}(y,E)L_xf_{r0}(x,E)]\,.
\eea
In case (ii) this function has a simple pole at $E=\a$.
In cases
 (i) and (iii) it is regular at $E=\a$ and can be calculated
as follows:
\bea
\nonumber
G_1(x,y,\a)\\  \label{G1lra}
=\left[\frac{\p}{\p E}
\frac{\T(y-x)L_x f_{l0}(x,E)L_yf_{r0}(y,E)+
\T(x-y)L_yf_{l0}(y,E)L_xf_{r0}(x,E)}{W_0}\right]_{E=\a }
\eea
\end{Co}
To prove these formulae we substitute $G_0$ as given in (\ref{Gteta})
into (\ref{G1}) and (\ref{G1a}).
Taking the derivative of the theta functions in (\ref{Gteta})
gives rise to the Dirac delta function which cancels out the delta function
present in (\ref{G1}).
Formula (\ref{G1lr}) is clearly valid since
 the SUSY transformations necessarily preserve the
boundary conditions for all $E$ except perhaps for $E=\a$. This implies
that $f_{l1}=Lf_{l0}$ vanishes at $x=a$ and $f_{r1}=Lf_{r0}$
vanishes at $x=b$. The denominator in
(\ref{G1lr}) is just the Wronskian of $f_{r1}$ and $f_{l1}$
and may be given in the form $W(f_{r1},f_{l1})=(E-\a)W(f_{r0},f_{l0})=(E-\a)W_0$.
\hfill$\square$

\section{Trace formulae}

The trace of the Green function defined as \cite{Morse}
$\int_a^bG(x,x,E) dx$ is usually divergent if the system has a continuous
spectrum. It is remarkable that the trace of the difference
$G_0(x,x,E)-G_1(x,x,E)$ is a finite quantity which may or may
not be equal to zero. In some cases this fact may be
explained by another remarkable property. It may happen that the
difference of infinite normalizations (they diverge as
$\delta(x-y)$ when $y\to x$) of the continuous spectrum
eigenfunctions of the two SUSY partners is a finite quantity.

\begin{Th}
Let $f_{l0}(x,E)$ and $f_{r0}(x,E)$ be solutions of the \Sc equation for
$h_0$ satisfying the zero boundary conditions at the left and
right bound of the interval $(a,b)$ respectively,
 and
\be\label{f1lr}
 f_{l1}(x,E)=Lf_{l0}(x,E)\qquad\qquad
f_{r1}(x,E)=Lf_{r0}(x,E)
\ee
be similar solutions for $h_1$
related with $h_0$ by a SUSY transformation with $\a$ being the
factorization constant.
Let $W_0$ be the Wronskian of $f_{r0}$ and $f_{l0}$,
$W_0=W(f_{r0},f_{l0})$.
Then
\be\label{traces}
\int_a^b[G_0(x,x,E)-G_1(x,x,E)]dx=\frac{Q(E)}{W_0(E-\a)}
\ee
where $Q(E)$ can be
calculated by one of the following formulae:
\bea\label{FF1}
&Q(E)
=
(f_{r0}f_{l1})_{x=b}-(f_{r0}f_{l1})_{x=a}=
(f_{l0}f_{r1})_{x=b}-(f_{l0}f_{r1})_{x=a}\\
&=-W_0+(f_{l0}f_{r1})_{x=b}-(f_{r0}f_{l1})_{x=a}
=
W_0+(f_{r0}f_{l1})_{x=b}-(f_{l0}f_{r1})_{x=a}\,.\label{FF2}
\eea
\end{Th}

{\it Proof.}
From Corollary 1 it follows that
$G_1(x,x,E)=\frac{1}{W_0(E-\a)}Lf_{l0}(x)Lf_{r0}(x)$.
While integrating this expression over the interval $(a,b)$ one
can transfer the derivative present in $L$ either from $f_{l0}$ to
$f_{r0}$ or from $f_{r0}$ to $f_{l0}$ which leads to one of the following
integrands $f_{l0}(x)L^+Lf_{r0}(x)$ or  $f_{r0}(x)L^+Lf_{l0}(x)$.
In both cases the factorization property may be used to reduce the
integrand to
$(E-\a)f_{l0}(x)f_{r0}(x)$. Thus we arrive at the relation
\be\label{th2}
\int_a^bG_1(x,x)dx=\frac 1{W_0}\int_a^b{f_{l0}(x)f_{r0}(x)}dx-
\frac{Q(E)}{W_0(E-\a)}
\ee
where $Q(E)$ is given by (\ref{FF1}).
To prove (\ref{FF2}) it is sufficient to notice that
\be\label{f01}
f_{l0}(x,E)f_{r1}(x,E)-f_{r0}(x,E)f_{l1}(x,E)=W_0
\ee
which is a consequence of (\ref{f1lr}).
The identification of the integrand on the
right hand side of (\ref{th2}) as $W_0G_0(x,x)$ then leads to the result
given in (\ref{traces}).
 \hfill$\square$

Using the first of equalities (\ref{FF1}) one can
rewrite (\ref{traces}) as follows:
\be\label{G01E0}
\int_a^b[G_0(x,x,E)-G_1(x,x,E)]dx=\frac{1}{\a-E}+
\frac{(f_{l0}f_{r1})_{x=b}-(f_{r0}f_{l1})_{x=a}}{W_0(E-\a)}
\ee
Now for the case (i) where $\a=E_0$ if we compare this result
with
the corresponding difference which
can be obtained directly from the expressions for $G_0$ given by (\ref{G0})
and for $G_1$ given by (\ref{G1i})
 the following feature may be noted:
 the
first term on the right hand side of (\ref{G01E0}) arises
from the contribution to Green functions from the discrete spectra
and the second term, which as we show below may be different of zero,
 is due to the presence of the continuous spectra. Just this
 contribution was neglected in \cite{Sukumar}. So, Theorem 2
 presents a generalization of the result obtained in
 \cite{Sukumar} to the case where a continuous spectrum may be
 present.
 As an application of this theorem we are going to consider two
 particular cases of scattering potentials defined both on the
 whole real line and on a semiaxis.

\begin{Co}
 If $h_0$ is a scattering Hamiltonian with the
 potential $V_0$ satisfying for the spectral problem on the whole
 line the condition
 \[
\int_{-\infty}^\infty (1+|x|)|V_0(x)|dx<\infty
 \]
 then for $E\ne \a$, $\mbox{Im}\sqrt{E}>0$ the following equality
 \be\label{trscat}
\int_{-\infty}^\infty [G_0(x,x,E)-G_1(x,x,E)]dx=
\frac{\d}{\k^2+ia\k}-\frac{\d}{\k^2+a^2}
\ee
holds, where $E=\k^2$, $\a=-a^2$;
$\d=1$ for the case (i), $\d=-1$ for the case (ii) and
 $\d=0$ for the case (iii).
\end{Co}

{\it Proof.}
The statement readily follows from the fact that any scattering
potential has a pair of solutions (Jost solutions, see e.g.
\cite{Levitan})
with the following asymptotics
at the right infinity
\[
f_{l,r}(x,E)\to e^{\mp i\k x}\qquad\quad E=\k^2
\qquad \quad \mbox{Im}\k>0
\qquad \quad x\to \infty
\]
and similar asymptotics at the left infinity and
the use of an appropriate part of equalities (\ref{FF1})
and (\ref{FF2}). The Wronskian $W_0$ for Jost solutions can easily
be calculated, $W_0=-2i\k$.
\hfill$\square$

So, we see that despite the fact that
 for both $h_0$ and $h_1$ the
continuous spectrum eigenfunctions are normalized to the Dirac
delta function, ({\it i.e.}) that in both cases they have equal infinite
norms, the difference of these infinities is a finite non-zero quantity in
cases (i) and (ii) and it is zero in case (iii).

For instance in case (ii) the
following equality arises:
\be\label{Ptrans}
\int_{-\infty}^\infty \frac{P(k)dk}{k^2-E}=R(E)
\qquad\quad
R(E)=\frac{-1}{\k^2+ia\k}\quad\quad E=\k^2\quad\quad \a=-a^2
\ee
where
\be\label{Pk}
P(k)=\int_{-\infty}^\infty [\,|\psi_k(x)|^2-|\chi_k(x)|^2]dx
\,.
\ee
Equation (\ref{Ptrans}) may be
reduced to the Stieltjes transform and the function $P$ may be
found by the Stieltjes inversion formula (see e.g.
\cite{Levitan0}).
To establish this we first notice that the integral on the
left hand side of (\ref{Ptrans}) is different from zero only if
$P(k)$ is an even function which we assume to be the case.
Therefore it can be considered only
for positive $k$s and we can let $k^2=\lb$. So, (\ref{Ptrans})
takes the form
\[
\int_{-\infty}^\infty\frac{d\rho(\lb)}{\lb-E}=R(E)
\]
where the measure $\rho(\lb)$ is continuous for $\lb>0$,
$d\rho(\lb)=\frac{1}{\sqrt{\lb}} P(\lb)d\lb$ and such that for negative
$\lb$s the integral is zero.
Now the Stieltjes inversion formula
yields
\[
\frac{P(\lb)}{\sqrt{\lb}}=\frac{\mbox{sign}\tau}{2\pi i}
\lim_{\tau\to0}[R(E)-R(\ov E)]\qquad\qquad E=\lb+i\tau
\]
where the bar over $E$ denotes the complex conjugate to $E$.
Note that because of the condition $\mbox{Im}\sqrt{E}>0$ the square
root of $E$ has different signs for $E$ in the upper and lower halves of
the complex $E$-plane. Therefore
the function $R(E)$ has a cut along the real
axis and the jump across this cut defines the function
$P(\lb)$.
After a simple calculation one gets
\be\label{Pla}
P(\lb)=a\pi^{-1}(\lb^2+a^2)^{-1}\,.
\ee
It must be noted that
in the present case the interchange of
the
integrals over the space variable $x$
taken in the difference of (\ref{G0})
and
(\ref{G1i}) at $y=x$
 with the integral over
the momentum $k$ is justified.

\begin{Co}
 If $h_0$ is a scattering Hamiltonian with the
 potential $V_0$ for the spectral problem on a half
 line satisfying the condition
 \[
\int_{0}^\infty x|V_0(x)|dx<\infty
 \]
 then there exist only two kinds of SUSY transformations keeping the zero boundary
 condition at the origin.
If $h_0$ has the discrete spectrum its ground state $\psi_0$ may be
deleted, ($u=\psi_0$, case (i)) and there is a possibility to keep the
spectrum unchanged (case (iii)).
The last possibility may be realized with $u(x)=f_l(x,E)$,
 $E<0$ where $f_l$ is such that $f_l(0,E)=0$.
 In this case the
 following trace formula is valid:
 \be\label{Semiscat}
\int_{0}^\infty [G_0(x,x,E)-G_1(x,x,E)]dx=
\frac{\d_2}{2(\k^2-ia\k)}-\frac{\d_1}{\k^2+a^2}
\ee
where $E=\k^2$ and $\a=-a^2$; for the case (i) $\d_1=\d_2=1$ and for the case (iii)
$\d_1=0$, $\d_2=-1$.
\end{Co}

{\it Proof.}. The proof is based on the fact that for such
potentials the left solution goes to zero like $x$ when $x\to0$
and the right solution has the asymptotics
$f_{r0}\sim\exp(i\k x)$, $E=\k^2$, $\mbox{Im}k>0$ (see e.g.
\cite{Faddeev}). It follows from here that
when $E$ is not a spectral point
 the left solution has a
growing asymptotics at infinity,
$f_{0l}\sim -\frac{W_0}{2i\k}\exp(-i\k x)$, $x\to\infty$, where $W_0$ is the
Wronskian of $f_{r0}$ and $f_{l0}$. It is not possible to create a
new bound state in this case since the \Sc equation with such a
potential has no solutions going to infinity as $x$ approaches
the origin.\hfill$\square$

{\em Example 1.} Free motion on the line, $V_0(x)=0$, $x\in\Bbb R$\\
The Green function is
\[
G_0(x,y,E)=\frac{i}{2\k}e^{i\k |x-y|}\qquad\qquad
\mbox{Im}\k>0\quad\quad E=\k^2\,.
\]
The choice $u=\cosh(ax)$, $\a=-a^2$ leads to the one soliton potential
$V_1=-2a^2\mbox{sech}^2(ax)$ with the Green function
\[
G_1(x,y,E)=if_\k(x)f_{-\k}(y)/[2\k(\k^2+a^2)]\qquad\qquad x\le y
\]
where
$f_\k(x)=\exp(-i\k x)(i\k+a\tanh ax)$, which clearly has a pole at
the ground state energy $E=-a^2$. The residue at the pole is
$\vfi_0(x)\vfi_0(y)$ where $\vfi_0(x)=\sqrt{a/2}\,\mbox{sech}(ax)$ which
is just the ground state of the one-soliton potential.

Continuous spectrum eigenfunctions of $h_0$,
$\psi_k(x)=1/\sqrt{2\pi}\exp{(ikx)}$ are transformed into continuous
spectrum eigenfunctions for $h_1$,
$\xi_k(x)=[-ik+\tanh(ax)]\exp{(ikx)}/\sqrt{2\pi(k^2+a^2)}$.
The direct calculation of the function $P(k)$ given in (\ref{Pk})
gives exactly the result (\ref{Pla}).

{\em Example 2.} Free motion on a half-line with zero
angular momentum, $V_0(x)=0$, $x\in\Bbb R^+$.\\
The Green function is
\[
G_0(x,y,E)=\frac 1\k\sin(\k x)\exp(i\k y)\qquad\quad
E=\k^2\quad\quad \mbox{Im}\k>0
\quad\quad x\le y\,.
\]
It must be noted that the use of the transformation function
$u=\cosh(ax)$ gives the same one-soliton potential
as in the previous example
which when
considered on a half-line has only a continuous spectrum but its
continuous spectrum eigenfunctions cannot be obtained by applying
the transformation operator $L$ to the free particle
eigenfunctions since it does not preserve the zero boundary
condition at the origin. Another choice $u=\sinh(ax)$ creates the
potential $V_1=2a^2\mbox{csch}^2(ax)$ which is singular at the origin and
 has only a continuous spectrum and which may be obtained with
the help of the SUSY transformation from the free particle
Hamiltonian. Its Green function is
\[
G_1(x,y,E)=\frac{ie^{i\k y}}{\k(\k^2+a^2)}
[\k+ia\mbox{cth}(ay)][\k\cos(\k x)-a\mbox{cth}(ax)\sin(\k x)]
\]
\[
E=\k^2\qquad \qquad\mbox{Im}\k>0\,.
\]
It is not difficult to see that despite the presence of the
denominator, $G_1$ is regular for all $\k\ne0$
including the point $\k=ia$ and discontinuous along
the positive real axis in the complex $E$-plane. The jump across
this cut is proportional to the product of continuous spectrum
eigenfunctions of $h_1$ which are given by
\[
\xi_k=\sqrt{\frac{2}{\pi(k^2+a^2)}}[-k\cos(k x)+
a\mbox{cth}(ax)\sin(k x)]\,.
\]
In contrast to the previous example the difference  $\psi_k^2-\xi_k^2$ now
 oscillates when $x\to\infty$,
the Riemann integral of this difference over the space variable is
divergent
 and the interchange of the
integrals over $k$ and $x$ is impossible.
Nevertheless, if one assumes that the improper integral over $x$ is
a limit of a proper integral
one can write
\be\label{P}
\lim_{A\to\infty}\int_{0}^\infty \frac{P(k,A)dk}{k^2-E}=R(E)
\qquad\qquad
R(E)=\frac{1}{2(ia\k-\k^2)}
\ee
$
\mbox{where }
E=\k^2,\ \a=-a^2 \mbox{ and }
P(k,A)=\displaystyle{\int}_{0}^A
[\,\psi_k^2(x)-\chi_k^2(x)]dx\ .
$
\\
In our example the function $P(k,A)$ is given by
\[
P(K,A)=\frac{2a\mbox{cth}(aA)\sin^2(kA)-k\sin(2Ak)}{\pi(k^2+a^2)}
\]
which when substituted into the left hand side of (\ref{P}) gives
exactly the function $R(E)$.
So this example shows that in contrast to the previous example
where the difference of normalisations is a finite quantity,
here this value is undetermined.
Nevertheless, the contribution to the trace of the difference
$G_0-G_1$ from the continuous spectra of $h_0$ and $h_1$ is well
defined.

\section{Conclusions}

In this paper we have studied the relation between the Green functions
corresponding to two Hamiltonians which are SUSY partners. We have shown
that it is possible to establish a relation between the traces of the
Green functions for the two partner Hamiltonians for the cases of
deletion of the ground state, the addition of a new ground state and when
the two Hamiltonians are isospectral. The formulae derived in this paper
are valid for the general case of Hamiltonians having both discrete
and continuous spectra. Our results show that when a continuous spectrum
is present, each of the traces of the Green functions for the SUSY partners
may diverge but the difference between the traces can be finite. We have
illustrated our results by considering the case of the free motion on the
full line and the case of the free motion of a particle with zero angular
momentum on the half-line.

Finally we would like to note that the difference of the traces
of the Green functions
of the two SUSY partner Hamiltonians appears as the trace (actually
super-trace) of the Green function of the supersymmetric \Sc
equation (supersymmetric Green function). Thus, our results reveal
the possibility of divergence of the component traces of the
supersymmetric Green function while its super-trace remains
finite.

\section*{Acknowledgments}

The work of BFS is partially supported by the
President Grant of Russia 1743.2003.2
 and the Spanish MCYT and European FEDER grant BFM2002-03773.

\end{document}